\documentclass[conference,dvipsnames]{IEEEtran}
\IEEEoverridecommandlockouts
\usepackage{cite}
\usepackage{amsmath,amssymb,amsfonts}
\usepackage{algorithmic}
\usepackage{graphicx}
\usepackage{textcomp}
\usepackage{xcolor}
\usepackage{xargs}
\usepackage{color,listings}
\usepackage[T1]{fontenc}
\usepackage[colorinlistoftodos,prependcaption,textsize=scriptsize]{todonotes}
\def\BibTeX{{\rm B\kern-.05em{\sc i\kern-.025em b}\kern-.08em
    T\kern-.1667em\lower.7ex\hbox{E}\kern-.125emX}}

\newcommandx{\rocky}[2][1=]{\todo[linecolor=Blue,backgroundcolor=Blue!25,bordercolor=Blue,#1]{\textbf{Rocky comments: }#2}}
\newcommandx{\mitra}[2][1=]{\todo[linecolor=OliveGreen,backgroundcolor=OliveGreen!25,bordercolor=OliveGreen,#1]{\textbf{Mitra comments: }#2}}
\newcommandx{\pragyan}[2][1=]{\todo[linecolor=Red,backgroundcolor=Yellow!25,bordercolor=Red,#1]{\textbf{Pragyan comments: }#2}}
\newcommandx{\tdb}[2][1=]{\todo[linecolor=Mulberry,backgroundcolor=Orchid!25,bordercolor=Mulberry,#1]{\textbf{Travis comments: }#2}}
\newcommandx{\sepideh}[2][1=]{\todo[linecolor=CadetBlue,backgroundcolor=CadetBlue!25,bordercolor=CadetBlue,#1]{\textbf{Sepideh comments: }#2}}
\newcommandx{\personF}[2][1=]{\todo[linecolor=Bittersweet,backgroundcolor=Bittersweet!25,bordercolor=Bittersweet,#1]{\textbf{PersonF comments: }#2}}
\newcommandx{\personG}[2][1=]{\todo[linecolor=RoyalBlue,backgroundcolor=ProcessBlue!25,bordercolor=RoyalBlue,#1]{\textbf{PersonG comments: }#2}}

\newcommandx{\citeme}[2][1=]{\todo[linecolor=red,backgroundcolor=red!25,bordercolor=red,#1]{\textbf{CITE} #2}}
\newcommandx{\fillin}[2][1=]{\todo[linecolor=red,backgroundcolor=red!25,bordercolor=red,#1]{\textbf{FILL IN} #2}}

\def\BibTeX{{\rm B\kern-.05em{\sc i\kern-.025em b}\kern-.08em
    T\kern-.1667em\lower.7ex\hbox{E}\kern-.125emX}}
\begin{document}

\title{An Analysis of Automated Use Case Component Extraction from Scenarios using ChatGPT
}

\author{\IEEEauthorblockN{Pragyan K C\textsuperscript{1}, Rocky Slavin\textsuperscript{1}, Sepideh Ghanavati\textsuperscript{2}, Travis Breaux\textsuperscript{3}, Mitra Bokaei Hosseini\textsuperscript{1}}
\IEEEauthorblockA{\textsuperscript{1}University of Texas at San Antonio, San Antonio, TX, USA, \textsuperscript{2}University of Maine, Orono, ME, USA, 
\\\textsuperscript{3}Carnegie Mellon University, Pittsburgh, PA, USA\\
\textit{
[pragyan.kc, rocky.slavin, 	mitra.bokaeihosseini]@utsa.edu, sepideh.ghanavati@maine.edu, breaux@cs.cmu.edu}
}

}

\maketitle

\begin{abstract}
 
Mobile applications (apps)  are often developed by only a small number of developers with limited resources, especially in the early years of the app's development. In this setting, many requirements acquisition activities, such as interviews, are challenging or lower priority than development and release activities. Moreover, in this early period, requirements are frequently changing as mobile apps evolve to compete in the marketplace. As app development companies move to standardize their development processes, however, they will shift to documenting and analyzing requirements. One low-cost source of requirements post-deployment are user-authored scenarios describing how they interact with an app. We propose a method for extracting use case components from user-authored scenarios using large language models (LLMs). The method consists of a series of prompts that were developed to improve precision and recall on a ground truth dataset of 50 scenarios independently labeled with UC components. Our results reveal that LLMs require additional domain knowledge to extract UC components, and that refining prompts to include this knowledge improves the quality of the extracted UC components.

\end{abstract}


\section{Introduction}

The popularity and usage of mobile applications (apps) among users have encouraged the industry to grow throughout many domains and categories. Mobile app development is similar to the traditional software development life cycle (SDLC)~\cite{wasserman2012software}. 
However, mobile app requirements can be introduced through constant feedback by end-users of mobile apps as a heterogeneous crowd in different geographical locations. This, coupled with the fact that the requirements for mobile apps change rapidly and continuously~\cite{abad2017learn}, makes traditional requirement elicitation methods, such as interviews, workshops, and prototyping, challenging. Such methods can be successfully applied when stakeholders (including users) are within organizational reach~\cite{oriol2018fame}. Therefore, the competitive mobile app development ecosystem presents unique challenges for developers to continuously elicit and analyze remote end-users' requirements to develop new apps or evolve existing apps' functionality and quality. These challenges are further highlighted for small app-developing companies with limited resources and developers, which must address demanding development timelines to compete in the market.

In such an ecosystem, similar apps can be a good reference for small companies to elicit and reuse requirements. However, the app requirements are generally not publicly available due to companies' policies~\cite{wei2022towards}. As an alternative, 
past studies have suggested using users' feedback, including reviews and comments, to elicit functional requirements in mobile apps~\cite{oriol2018fame,dalpiaz2019re,di2016would,maalej2015bug}. Such reviews are publicly available through app markets and can be accessed without cost or much effort. 
However, reviews are short (on average 71 characters~\cite{genc2017systematic}), noisy, and only 35.1\% of them have been found to contain information that directly helps developers to improve their apps~\cite{chen2014ar}. Further, app reviews are written in natural language with little to no structure~\cite{rago2013uncovering,luisa2004market}. 

Given the shortcomings of eliciting requirements from app reviews, we propose using user-authored scenarios as a source of mobile app requirements. 
To generate such scenarios, we design and conduct a survey on app users. In this survey, users are instructed to describe their goal in using a specific screen in an app, the steps they take to accomplish the goal, and the information the app requires to achieve the goal. This approach enables small app-developing companies to identify a sustainable and low-cost source of requirements using similar apps. To develop a new app in a specific category (e.g., lifestyle), developers can solicit users that use apps in the same category to provide scenarios for a small payment. In addition, developers can recruit users to evolve their existing app through access to their contact information or in-app notifications. In this case, developers can choose specific screens from their apps that are tied to new feature introductions, changes, or app maintenance. We also design a study to evaluate the quality of user-authored scenarios in conveying the goals, interactions between actors, and information types required to achieve the goal. 

To further assist developers, 
we explore whether use case (UC) components can be extracted from the scenarios. UCs are effective structured techniques for presenting functional requirements~\cite{cockburn2001writing,fantechi2003applications} and have gained wide acceptance among requirements analysts, designers, and testers~\cite{el2012constructing}. Given user-authored scenarios as a source of requirements for a mobile app, an analyst can extract UC components (e.g., actors, goals, and interactions) by leveraging their own experience and skills, particularly those linguistic in nature~\cite{al2018use}. However, various challenges exist in extracting UC components. First, manually extracting UC components from requirements sources is time-consuming, error-prone~\cite{al2018use}, and does not scale considering the continuous changes in the requirements and the often small number of developers in the company. Second, developing UC components with low quality has potential consequences (e.g., cost of fixing defects at later stages of SDLC)~\cite{anda2001quality,el2006matching,el2006agaduc,el2010improving}, which companies should want to avoid. 

To address these challenges, we propose the use of existing Large-scale Language Models (LLMs) to automatically extract UC components from scenarios. For this reason, we select ChatGPT\footnote{https://openai.com/blog/chatgpt} as an instance of LLMs~\cite{vaswani2017attention} that recently has received significant attention. 
We design two studies to evaluate both the ability of ChatGPT to automatically extract UC components and the quality of the extracted components. Lastly, we explore the effects of prompt engineering on the quality of the extracted UC components through a case study. Prompts are instructions given to a
LLM to enforce rules, automate processes, and ensure specific
qualities (and quantities) of the generated output~\cite{white2023prompt}. 
To guide this analysis, we pose the following research questions.
\begin{enumerate}
\item[\textbf{RQ1}:] What is the quality of the scenarios authored by app users?
\item[\textbf{RQ2}:] To what extent is ChatGPT as a LLM able to extract UC components from scenarios?
\item[\textbf{RQ3}:] What are the defects in the UC components extracted using ChatGPT?
\item[\textbf{RQ4}:] What are the effects of prompt engineering on the quality of the UC components extracted using ChatGPT?
\end{enumerate}


The contributions of this paper are as follows: (1) a 50-scenario corpus as a source of mobile app requirements; (2) a data set of labeled scenarios with UC components; (3) empirical evaluation of ChatGPT's performance in extracting UC components (4) an analysis of the quality of the extracted UC components; and (5) an exploratory study of prompt engineering in relation to UC component quality.

This paper is organized as follows. In Section~\ref{sec:relatedWork}, we discuss the background and related work. In Section~\ref{sec:approach}, we introduce our approach for collecting \& labeling the data. In Section~\ref{sec:evaluation}, we describe the evaluation method. In Section~\ref{sec:results}, we present the results, followed by discussion, threats, and conclusion \& future work in Sections~\ref{sec:discussion}, \ref{sec:threats}, and \ref{sec:conclusion}, respectively.


\section{Background and Related Work}\label{sec:relatedWork}

\subsection{App Reviews}
Mobile app companies must continuously elicit requirements to improve their apps' functionality and keep up with the competitive app development market. 
Users' feedback (i.e., reviews and comments) has been suggested as a source of requirements in apps~\cite{oriol2018fame,dalpiaz2019re,di2016would,maalej2015bug,yang2015identification}, 

as it allows end-users to communicate their needs for the developed app anytime and anywhere~\cite{oriol2018fame}. 
However, reviews are short (71 characters on average~\cite{genc2017systematic}) and noisy. Further, only 35.1\% of reviews contain information that directly helps developers to improve their apps~\cite{chen2014ar}. Therefore, any approach to eliciting requirements must use filtering techniques and content aggregation~\cite{chen2014ar,pagano2013user,panichella2015can,oriol2018fame,harman2012app,iacob2013retrieving}. 
Even in the event of access to high-quality (i.e., clear, unambiguous, and complete) app reviews, 
requirements analysts still have to deal with the fact that they have little or no structure as they are written in natural language~\cite{rago2013uncovering,luisa2004market}. 
While this facilitates convenient communication by users, processing text written in natural language can be time-consuming and error-prone for the analyst. Finally, studies that try to elicit requirements from app reviews are difficult to replicate, with measured recall as low as 0.34~\cite{dkabrowski2020mining}. 

\subsection{User Scenarios}

Given the shortcomings of eliciting requirements from app reviews, we propose that mobile app developers use scenarios, where users describe their goal and a series of steps to interact with the app and accomplish the goal. 
Unlike user stories, scenarios are more detailed and include motivation and goals. In effect, scenarios are instances of experience with a system captured from users~\cite{sutcliffe1998scenario}. 
Scenario-based approaches are commonly used in requirements engineering to elicit, analyze, and validate requirements~\cite{alexander2005scenarios,anton1998representational,weidenhaupt1998scenarios}. For example, 
Hibshi et al. elicit security requirements from scenarios drawn upon specific examples from users' experience or background knowledge~\cite{hibshi2021systemic}. 
Shen and Breaux collect user-authored scenarios in the
directory service domain and analyze them to extract a domain
knowledge using typed dependency parsing and word embeddings~\cite{shen2022domain}. Scenarios improve the alignment of requirements with user needs and can discover requirements unforeseen by system analysts~\cite{sutcliffe1998scenario, potts1994inquiry}. The collection and analysis of scenarios in various app domains are used to construct formal domain models or knowledge bases that facilitate activities in the SDLC (e.g., maintenance phase, where scenarios can be used to add or improve functionalities and features)~\cite{sutcliffe1998scenario}. 
Further, scenarios can challenge the model assumptions and, therefore, be used to validate formal models and specifications~\cite{sutcliffe1998scenario,johnson1988computer}.

\subsection{Use Cases}
A \textit{Use Case} (UC) is a text document that captures the behavior of a system as it responds to a request from an individual actor~\cite{cockburn2001writing}. 
The idea of UCs to describe functional requirements was introduced in 1986 by Ivar Jacobson~\cite{jacobson1995use}, a primary contributor to the Unified Modeling Language (UML) and the Unified Process (UP)~\cite{larman1998applying}, and has since been used as a primary tool for requirements specification.
Since UCs describe interactions across the system's boundary; 
they can be an effective means for identifying data practices, highlighting the importance of UCs in software traceability~\cite{insfran2002requirements}.

There are various guidelines in the literature for modeling UCs~\cite{bittner2003use,ouergaard2005use}.
Cockburn presents guidelines for writing UCs using natural language and provides several templates that are informally considered as industry standards~\cite{cockburn2001writing}. Further, Cockburn proposes a pass/fail test containing 28 questions to check the quality of UC components. 
Similarly, Anda et al. evaluate the quality of the UC models using a taxonomy of defects~\cite{anda2002towards}. 
El-Attar and Miller identify the quality attributes of the UC models through a literature review~\cite{el2012constructing}.

Notably, UCs are generally not created directly by app users. Instead, they must be extracted from the raw requirements, such as user stories or scenarios. This can be difficult due to the problems inherent to natural language requirements~\cite{al2018use}. 
Al-Haroob et al. introduce a tool that utilizes neural networks and NLP to identify actions and actors from requirement documents~\cite{al2018use}. 
Kundi et al. use FrameNet to enhance the understandability of the semantics in natural language requirements and extract the UCs~\cite{kundi2017use}. 
These approaches only extract actors and actions, while our approach considers additional UC components, including name, goal, user, system, external entity, data practices, and steps.

Ultimately, UCs are valuable for refining high-level requirements into lower-level specifications components~\cite{dalpiaz2020conceptualizing, insfran2002requirements,maiden2004model,yue2011systematic}.
However, the refinement of such UCs depends on the quality of the source requirements. 
To this end, this paper aims to improve the availability of UCs in mobile app development through their extraction from high-level user scenarios.


\subsection{Question Answering (QA) Models} 
\textit{QA} is a commonly used NLP technique for generating or exacting an answer to a question based on the provided context. Enhancement of QA tasks began with the development of Transformers~\cite{vaswani2017attention}, which also helped in the development of models such as BERT~\cite{devlin2018bert}, RoBERTa~\cite{liu2019roberta} and GPT~\cite{radford2018improving}.

The GPT architecture consists of a stack of transformer blocks, each of which has a self-attention mechanism that allows it to capture the relationships between different words in a sentence~\cite{radford2018improving}. 
 
The progression of the GPT~\cite{radford2018improving} model to GPT-2~\cite{radford2019language}, GPT-3~\cite{brown2020language}, and InstructGPT~\cite{ouyang2022training} has led to the development of ChatGPT that recently received great attention. 
ChatGPT can extend text input, answer questions, and have a conversation based on previous chat history. In addition, it can also be used for source code generation and completion, e.g., to suggest a fix of incorrect code~\cite{sobania2023analysis}. 

In this paper, we evaluate the use of ChatGPT in extracting UC components.

\subsection{Prompt Engineering}

A \textit{prompt} is a set of instructions provided to a LLM that customizes, enhances, or refines its capabilities~\cite{white2023prompt}. 
Prompts are crucial in setting the context for the LLM to tell what information is important and what needs to be the output~\cite{white2023prompt}. The task performance depends significantly on the quality of the prompt, and most effective prompts have been handcrafted by humans~\cite{zhou2022large}. 
White et al. introduce prompt patterns to systematically engineer different output and interaction goals when working with LLMs~\cite{white2023prompt}.
Lui et al. explore the prompt keywords and model hyperparameters that can help produce coherent outputs~\cite{liu2022design}. 
Zhou et al. propose an approach for automatic instruction generation and selection~\cite{zhou2022large}. 
In this paper, we explore the effect of prompt engineering on the quality of UC components extracted using ChatGPT.

\section{Approach}\label{sec:approach}

\begin{figure}
  \includegraphics[width=\columnwidth, keepaspectratio]{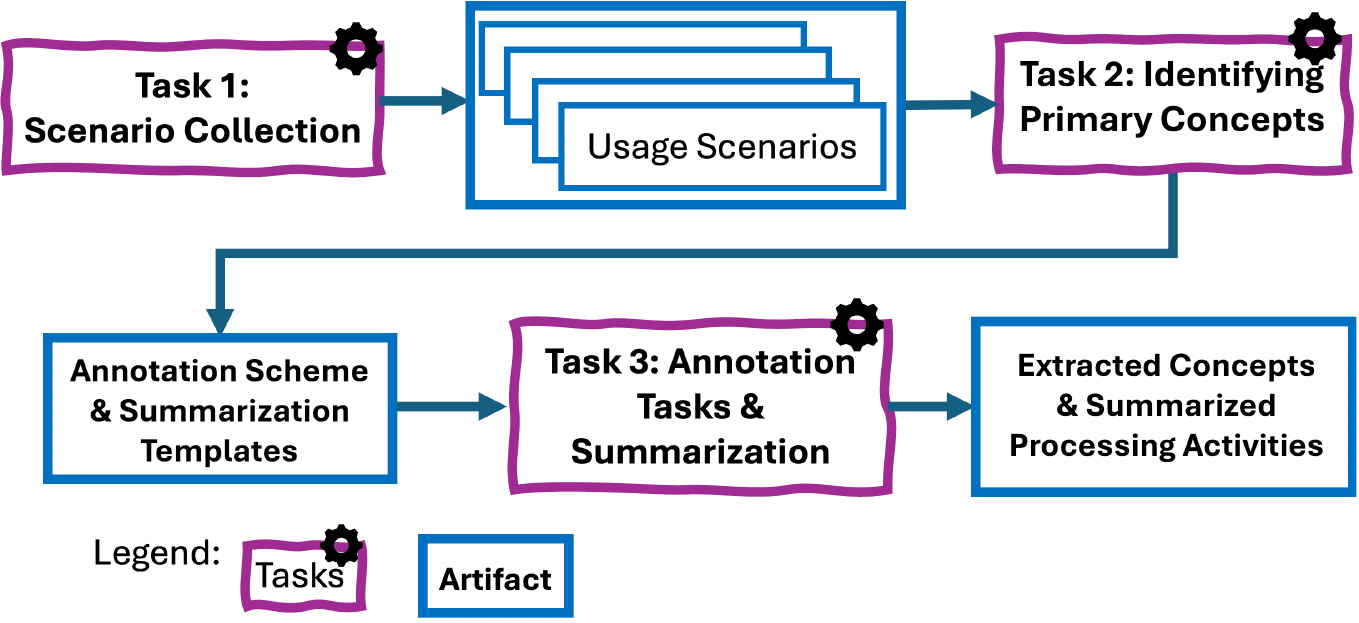}
   \centering
  \caption{The Overview of Scenario and UC Component Collection Approach}
    \label{fig:approach}
\end{figure}

Figure~\ref{fig:approach} illustrates our approach to collecting data required for designing our studies. First, we construct a user-authored scenario corpus. Second, we produce a ground truth by manually labeling the scenarios to extract UC components. 
Third, we utilize ChatGPT to automatically extract UC components from the scenarios. 
We now describe each step in detail.

\subsection{Scenario Corpus} 

To construct a corpus of usage scenarios from users, we design a survey and invite mobile app end-users to upload a screenshot from a mobile app, describe a usage scenario for the screenshot in at least 150 words, and answer some questions. The survey targets apps from various domains. 

\begin{figure}
{%
\setlength{\fboxsep}{5pt}%
\setlength{\fboxrule}{0.5pt}%
  \fbox{\includegraphics[width=0.93\columnwidth, keepaspectratio]{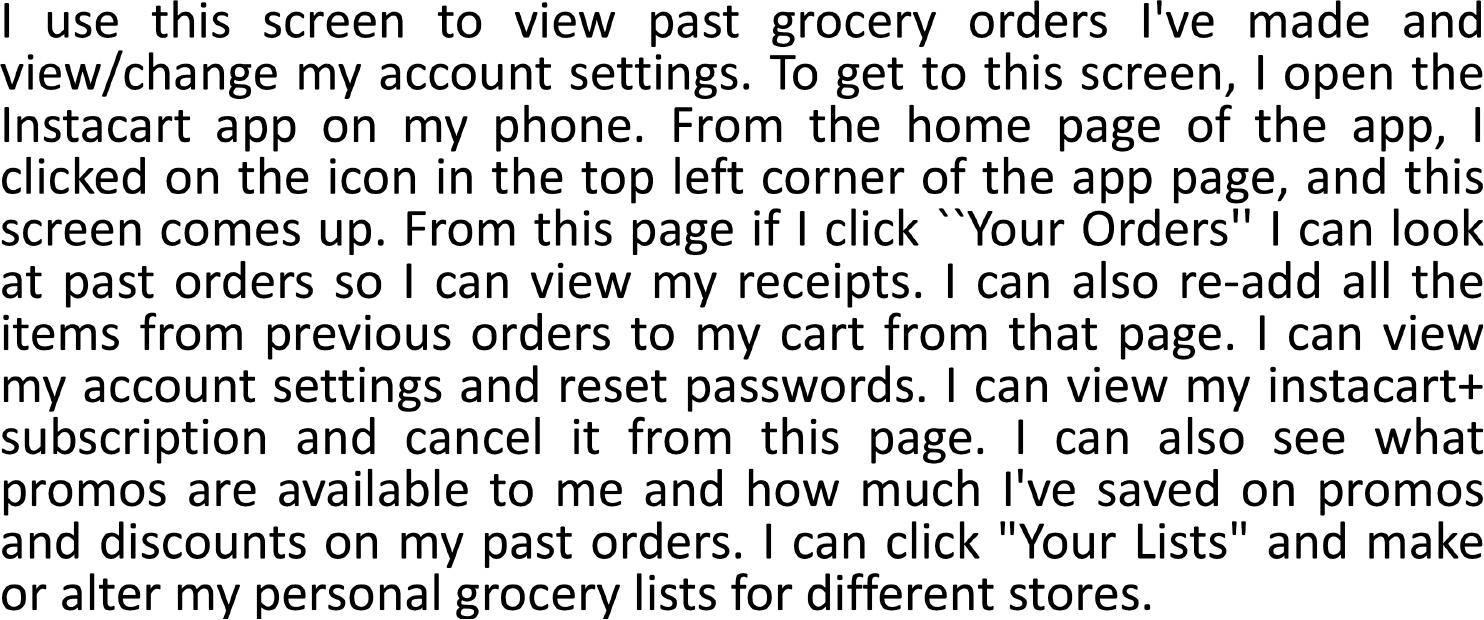}
  }%
  }%
   \centering
  \caption{A User-Authored Scenario Example}
    \label{fig:Example-Scenario}
\end{figure}

Survey participants are first asked to select a mobile app they use frequently and identify its URL from the Google Play or Apple App store. Second, participants are asked to identify three types of personal information that the app collects, uses, or shares for its functionality. Third, participants select a specific screen in the app and take a screenshot. Participants are instructed to select screens with the following properties: (a) emphasis on the core functionality of the apps; (b) not the app's homepage; (c) not the app's login page; and (d) not the app's settings page. We also provide the participants with examples of ideal and bad screenshots as further guidance in the instructions. To submit the screenshot from mobile phones, participants scan a QR code that navigates them to a webpage where they upload, redact any personal information, and submit the screenshot. Fourth, the submitted screenshot is loaded to the main survey page, where participants are asked to provide a title for the screen and write a usage scenario of at least 150 words that contains: (a) a description of the goal the user wants to achieve through the screen; (b) steps that they take to get to the screen on the app; and (c) the steps they take to achieve the goal once at the screen. The participants are also provided with an example of a scenario in the instructions.  
Figure~\ref{fig:Example-Scenario} illustrates an example of a user-authored scenario for viewing past grocery orders on the Instacart app.

The survey is published on the Amazon Mechanical Turk crowdsourcing platform. Workers who have completed 5,000 Human Intelligence Tasks (HITs), have an approval rating greater than 97\%, and are located in the United States are eligible to participate. Workers are compensated \$4.00 upon completion of the survey. As part of our protocol to protect human subjects, workers must provide informed consent before participating in the survey and the study is monitored by our Institutional Review Board (IRB).

\begin{table}[ht!]
\centering
\caption{Mobile App Scenario Frequency for Apple  and Google App Category}
\label{tab:appleGoogle_categories}
\begin{tabular}{|l|c|c|}
\hline
\textbf{App Category} & \textbf{Apple} & \textbf{Google} \\
\hline
Education & 2 & -\\
Finance &  7 & 1\\
Food \& Drink &  1 & 4\\
Health \& Fitness &  4 & 5\\
Reference &  1 & -\\
Shopping &  1 & 3\\
Social Networking & 1  & 5\\
Sports & 2 & 1\\
Travel &  1 & 2\\
Books \& Reference &  - & 1\\
Casual & - & 2\\
Entertainment & - & 1\\
Lifestyle & - & 1\\
Maps \& Navigation & - & 1\\
Music \& Audio & - & 1\\
Tools & - & 2\\
\hline
\textbf{Total} & \textbf{20} & \textbf{30} \\
\hline
\end{tabular}
\vspace{0.2cm}
\end{table}

In Table~\ref{tab:appleGoogle_categories}, we present the frequencies of apps per category described by the Apple App and Google Play stores, for the 50 scenarios authored by users. The table shows the diversity of scenarios we collected from various domains and strengthens the generalizability of our work. 
In practice, developers can elicit requirements for a new mobile app by first identifying similar app categories on the market (competitor analysis). They can then solicit users that utilize the apps from similar categories to author scenarios for a specific screenshot selected by the user for a small payment.

\subsection{Manual Extraction of UC Components} \label{sec:approach-manual}
In this subsection, we describe our approach for labeling the scenario corpus and extracting UC components manually. 

We adopt a simple template UC notation from Cockburn~\cite{cockburn2001writing} and consider seven UC components that can be extracted from a scenario, including \texttt{UC-Name}, \texttt{UC-Goal}, \texttt{UC-User}, \texttt{UC-System}, \texttt{UC-ExternalEntities} (\texttt{UC-ET}), \texttt{UC-DataPractices} (\texttt{UC-DPs}), and \texttt{UC-Steps}. The definitions and examples of these components are provided in Table~\ref{tab:UCcomponents}. We categorize the interactions between the actors (i.e., user, system, and external entities) as steps and data practices. Data practices are specific kinds of interactions that convey privacy requirements. We borrow the definition of privacy requirement from Breaux et al.~\cite{breaux2014eddy}, where a privacy requirement consists of actors with whom the data is shared, actions that are performed on the data, data elements on which actions are performed, and purposes for which data maybe be acted upon. 
This refined categorization of interactions between actors can help developers to assess the privacy risks associated with a data practice and further identify specific privacy patterns to mitigate risks when designing apps based on the requirements~\cite{hoepman2014privacy,colesky2016critical,colesky2016privacy}. 

\begin{table}
\caption {UC Components Definition} \label{tab:UCcomponents}
\center
\begin{tabular}{ |p{1.4cm}|p{6.5cm}| }
 \hline
  \textbf{UC \newline Component} & \textbf{Definition} \\
  \hline
  UC-Name & A name is a label that describes the purpose of the UC. Usually, VERB, NOUN, or a combination of VERB and NOUN is sufficient as the UC name, e.g., ``Order.''\\
  \hline
  UC-Goal & Describe the UC's goal, e.g., ``Ordering something successfully from McDonald's.''\\
  \hline
  UC-User & The primary user is the person using the mobile app and describes the scenario. Pronouns can be an example of the primary user in the scenario.\\
  \hline
  UC-System & The primary system involved in the UC. For example, ``McDonald's app'' or ``McDonald's'' are the primary system in a scenario.\\
  \hline
  UC-ET & Any other actor/system besides the primary user and the system mentioned in the scenario should be considered an external entity. For example, ``Google Pay'' is an external entity in the following sentence: ``I use Google Pay to pay for McDonald's order.''\\
  \hline
  UC-DPs & Any data practice entailing the collection, usage, and sharing of information types, e.g., ``app uses my location'' or ``I provide my address.''\\
  \hline
  UC-Steps & List of actions performed by the actors or the system, e.g., ``view available items'' or ``click the reorder button.''\\
\hline
\end{tabular}
\end{table}

To label the UC components in each scenario from the scenario corpus, we created a private labeling job on Amazon SageMaker Ground Truth. 
 We also created a private team, including four software engineering researchers who have privacy expertise (one faculty and three undergraduate students) to label the scenarios based on the definitions in Table~\ref{tab:UCcomponents}. The input file for this labeling task is published online\footnote{https://anonymous.4open.science/r/ArtifactsResearch-24FA} in JSON format. We set the number of distinct workers to perform the same labeling task as three. We measure the chance agreement between workers using Fleiss' Kappa\cite{fleiss1971measuring}. 

During this task, each worker individually analyzed five scenarios. Then, the workers met to discuss the scenarios and UC components, where they identified heuristics to clarify boundary and edge cases. As a result, workers developed heuristics that help increase the chance of agreement between them. Here we list such heuristics developed by the workers. 
\begin{itemize}
    \item H1: The goal is most likely in the first or second sentence, and it is preceded by phrases, such as ``I want to,'' ``my aim is,'' or ``I want to achieve.''
    \item H2: If the goal is vague or not otherwise obvious, workers still continue to label the scenario for other components. 
    \item H3: The goal that the user wants to achieve relates to the entire screen, and users may state sub-goals in terms of steps with pre-conditions or data practices.
    \item H4: Label every instance of the goal if the user has restated the goal in the scenario. 
    \item H5: The UC names (\texttt{UC-Name}) are mostly a sub-sequence of the goal (i.e., \texttt{UC-Goal}).
    \item H6: A phrase (e.g., ``this screen'') can be labeled with \texttt{UC-System} if the sentence holds its meaning after substituting the phrase with ``the system'' or ``app.'' 
    \item H7: The functions users can perform through the screen should be labeled as steps or data practices. 
    \item H8: Data practices involve the flow of potentially sensitive data. 
    \item H9: Steps and data practices are not part of the \texttt{UC-Goal}. 
\end{itemize}

\begin{table}
\caption {Kappa for Manual Labeling Task} \label{tab:kappa}
\center
\begin{tabular}{ |p{0.8cm}|p{0.8cm}|p{0.8cm}|p{0.8cm}|p{0.8cm}|p{0.8cm}|p{0.8cm}| }
 \hline
 UC-Name & UC-Goal & UC-User & UC-System & UC-ET & UC-DPs & UC-Steps\\
 \hline
 0.59\% & 0.67\% & 0.70\% & 0.68\% & 0.73\% & 0.70\% & 0.68\%\\
 \hline
\end{tabular}
\end{table}
Using the heuristics, the annotators reach Fleiss' Kappa values for each UC component presented in Table~\ref{tab:kappa}. Analyzing the kappa values, the workers have reached a moderate agreement for \texttt{UC-Name} and substantial agreement across the other six UC components~\cite{landis1977measurement}.  


\begin{table}[ht!]
\centering
\caption{UC Components Manually Labeled by Workers}
\label{tab:Example-Use-Case}
\begin{tabular}{|p{1.4cm}|p{6.5cm}| }
\hline
\textbf{UC \newline Component} & \textbf{Values} \\
\hline
UC-Name & [``''] \\
\hline
UC-Goal & [``view past grocery orders I've made'', ``view/change my account settings'', ``look at past orders'']\\
\hline
UC-User & [``I'']\\
\hline
UC-System & [``Instacart app'', ``the app'']\\
\hline
UC-ET & [``'']\\
\hline
UC-DPs & [``re-add all the items from previous orders to my cart'', ``view my receipts'', ``reset passwords'', ``make or alter my personal grocery lists'',``how much I've saved on promos and discounts on my past order'']\\
\hline
UC-Steps & [``clicked on the icon in the top left corner'', ``screen comes up'', ``click Your Orders'', ``view my account settings'', ``view my instacart+ subscription'', ``cancel it'', ``see what promos are available to me'', ``click Your Lists'']\\
\hline
\end{tabular}
\vspace{0.2cm}

\end{table}
Table~\ref{tab:Example-Use-Case} illustrates the UC components for the Instacart scenario (see Figure~\ref{fig:Example-Scenario}). This table presents the value for a UC component using a list of strings separated by commas. UC components with no values are represented as an empty list of strings (see \texttt{UC-ET} value in Table~\ref{tab:Example-Use-Case}).

\subsection{Automated Extraction of UC Components}

\begin{figure}
\centering
{%
\setlength{\fboxsep}{5pt}%
\setlength{\fboxrule}{0.5pt}%
  \fbox{\includegraphics[width=0.7\columnwidth, keepaspectratio]{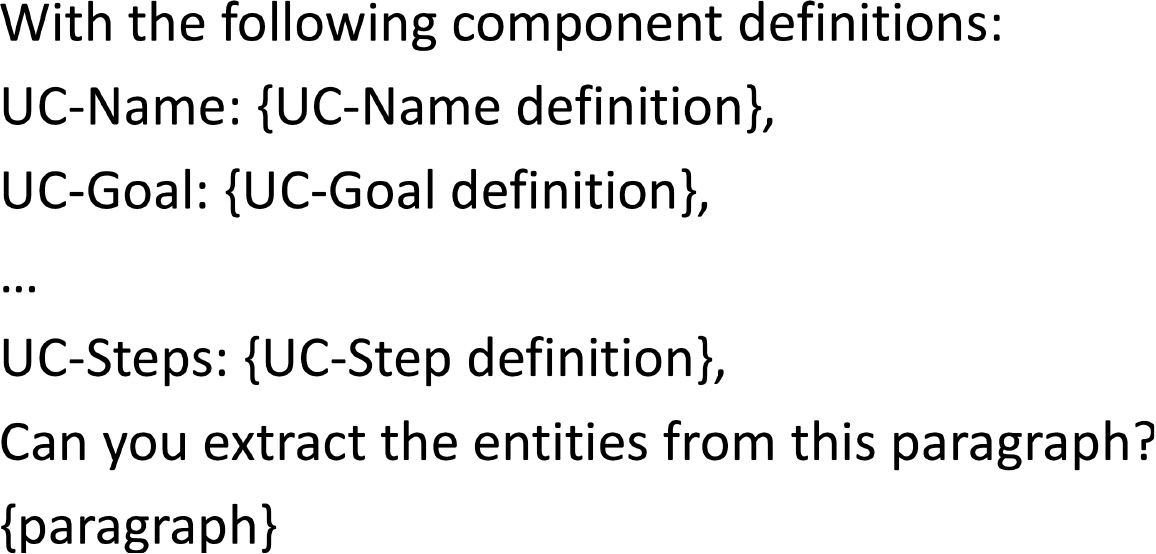}}
  }%
  \caption{Seed Prompt}
    \label{fig:seed-prompt}
\end{figure}

We now describe our approach for extracting UC components automatically using ChatGPT. 
To extract UC components from a scenario, we provide a seed prompt (Figure~\ref{fig:seed-prompt}) to ChatGPT. The curly brackets starting with ``UC-'' in this figure are replaced with the definitions of UC components (see Table~\ref{tab:UCcomponents}). By providing the definition of the UC components in separate lines in the seed prompt, we instruct ChatGPT to generate an answer with the same structure. Further, ``\{paragraph\}'' is replaced with a scenario from our corpus. We repeat this prompt once for each scenario in the corpus and record the results. 

Table~\ref{tab:Example-QA} illustrates an example of extracted UC components using our seed prompt for the scenario presented in Figure~\ref{fig:Example-Scenario}. In this table, each UC component is mapped to a comma-delimited list of strings.

\begin{table}[ht!]
\centering
\caption{UC Components Extracted from ChatGPT }
\label{tab:Example-QA}
\begin{tabular}{|p{2cm}|p{5.5cm}| }
\hline
\textbf{UC Component} & \textbf{Values} \\
\hline
UC-Name: & [``View Past Orders and Account Settings'']\\
\hline
UC-Goal: & [``To allow the user to view past orders and modify account settings'']\\
\hline
UC-User: & [``User'']\\
\hline
UC-System: & [``Instacart App'']\\
\hline
UC-ET: & [``'']\\
\hline
UC-DPs: & [``Collection'', ``Usage'', ``Sharing'']\\
\hline
UC-Steps: & [  ``Open Instacart app on phone'', ``Click on icon in top left corner'', ``Select Your Orders to view past orders and receipts'', ``Add items from previous orders to cart'', ``View account settings and reset passwords'', ``View and cancel Instacart+ subscription'', ``View available promos and discounts'', ``Click on Your Lists to create or modify personal grocery lists'']\\
\hline
\end{tabular}
\vspace{0.2cm}
\end{table}

\section{Evaluation}\label{sec:evaluation}




In this section, we describe the design of four studies to evaluate our research questions. 
To address RQ1, we analyze the quality of user-authored scenarios in study 1. Study 2 addresses RQ2, comparing the manually labeled ground truth UC components with the automatically extracted UC components from ChatGPT using statistical metrics. 
We design study 3 to address RQ3, which utilizes a checklist to analyze the quality of the UC components automatically extracted by ChatGPT. 
Finally, we design an exploratory case study to address RQ4 and explore whether prompt engineering affects the quality of the extracted UC components.

\subsection{Study 1 Design}
We now describe our approach to evaluate the quality of the 50 user-authored scenarios in our corpus and address RQ1. As discussed in Section~\ref{sec:approach}, we collect user-authored scenarios through a survey in which participants are instructed to submit an app screenshot and provide a minimum of 150 words describing the goal they want to achieve through the screen, the steps they take to accomplish the goal, and the information types the app needs to achieve the goal. To evaluate the quality of the scenarios and analyze whether users followed the survey instruction, we utilize the manually labeled scenarios by workers in Section~\ref{sec:approach-manual}. We specifically check if a scenario from the corpus contains manually labeled \texttt{UC-Goal}, \texttt{UC-Steps}, and \texttt{UC-DPs}. These three components align with the user instructions in the scenario collection survey.

\subsection{Study 2 Design}

The evaluation of QA models relies on human-annotated data sets of question-answer pairs~\cite{risch-etal-2021-semantic}. Given a question, the ground truth is compared to the answer predicted by a model using different metrics. 
In this section, we describe our approach for comparing the manually labeled UC components (i.e., ground truth) with the ChatGPT automatically extracted UC components (i.e., prediction) to address RQ2. For each scenario, we construct two dictionaries, one containing the UC components in the ground truth and the other for the components in the prediction. 
A dictionary contains keys representing the UC components, and each key is mapped to a list of strings associated with that specific UC component. We refer to each string as an element of the list. For example, Tables~\ref{tab:Example-QA} and ~\ref{tab:Example-Use-Case} show the lists of strings mapped to each UC component in the ground truth and prediction for the scenario presented in Figure~\ref{fig:Example-Scenario}. 

We compare the ground truth with the predicted UC components using three QA metrics, including the exact match (EM), F1-score, and semantic similarity (SM). 
EM is a binary metric that checks whether the ChatGPT prediction matches exactly the ground truth. This metric works well for factual answers, where the answer to a question is the name of a person or location. 
The F1-score relies on precision and recall, where precision is the relative number of tokens in the prediction that are also in the ground truth, and recall is the relative number of tokens in the ground truth that are also in the prediction. For true positives (TP), false positives (FP), and false negatives (FN), the precision is equal to $TP / (TP + FP)$; recall is equal to $TP / (TP + FN)$; and F1-score is the harmonic mean and equal to $(precision \times recall) / (precision + recall)$. 
To calculate the precision and recall for a UC component, we follow these steps: (1) we concatenate all the strings in the list mapped to the UC component in the ground truth into one string delimited with comma, and tokenize the string, yielding a new list of tokens $GT_{tokens}$; (2) we concatenate all the strings in the list mapped to the same UC component in the prediction (i.e., extracted components from ChatGPT) into one string delimited with space and tokenize the string, yielding a new list of tokens $P_{tokens}$; (3) for both lists $GT_{tokens}$ and $P_{tokens}$, we remove the redundant tokens, yielding two sets containing unique tokens $GT_{uTokens}$ and $P_{uTokens}$; (4) to calculate TPs, we identify the number of tokens shared words between $GT_{uTokens}$ and $P_{uTokens}$; (5) to calculate FPs, we count the number of tokens appearing in $P_{uTokens}$ that do not exist in the $GT_{uTokens}$; (6) to calculate FNs, we count the number tokens appearing in $GT_{uTokens}$ that do not exist in $P_{uTokens}$. 
We repeat these steps for all the UC components in the scenario corpus and calculate precision, recall, and F1-score for each scenario. The overall F1-score of a UC component is the average score for all 50 scenarios. 

We also analyze whether pre-processing, ground truth, and prediction would improve the F1-score. To pre-process the strings, we remove the punctuation and stop words and turn the tokens into their lemmas. We repeat steps 1-6 for the pre-processed data and calculate the precision, recall, and F1-score. 

The EM metric and F1-score are purely lexical and rely on string matching. Therefore, they will not account for the contextual similarity between the ground truth and prediction. 
To address this problem, we use cosine similarity as our SM metric to compare the ground truth and prediction. This metric requires vector representation (i.e., embedding) of strings for comparison. 
To this end, we create vector embeddings for the strings in ground truth and prediction as follows: (1) we concatenate all the strings in the list mapped to the UC component in the ground truth into one string delimited with commas, yielding a new string $GT_{string}$; (2) we concatenate all the strings in the list mapped to the same UC component in the prediction into one string delimited with commas, yielding string $P_{string}$; (3) We embed $GT_{string}$ and $P_{string}$ into two 1,024 dimension vectors $GT_{vector}$ and $P_{vector}$ using \texttt{stsb-roberta-large} model~\cite{liu2019roberta} from the Huggingface\footnote{https://huggingface.co/sentence-transformers/stsb-roberta-large} \texttt{Sentence-Transformer} library.

Cosine similarity is the cosine of the angle between the two provided vector representations. The cosine similarity value ranges from -1 to 1, where -1 is unrelated, and 1 is identical.  
We use the \texttt{pytorch\_cos\_sim} function from the sentence-transformer library to compute the semantic similarity between the two vectors $GT_{vector}$ and $P_{vector}$. We report the average SM for each component in 50 scenarios.

\subsection{Study 3 Design}
\label{sec:evaluation-study3}

\begin{table*}
\centering
\caption{Quality Assessment Questionnaire for UC Components}
\label{tab:questionnaire}
\begin{tabular}{|p{1.4cm}|p{15.5cm}| }
\hline
\textbf{UC \newline Component} & \textbf{Questions} \\
\hline
 UC-Actor & Q1. Are there any actors that are not identified in the extracted UC-User, UC-System, or UC-ET components?\\
 & Q2. Are there any incorrect actors in the extracted UC-User, UC-System, or UC-ET components?\\
& Q3. Considering the extracted UC-User, UC-System, UC-ET, and  UC-steps, are there actors not involved in at least one of the steps?\\
& Q4. Considering the extracted UC-User, UC-System, UC-ET, and  UC-DPs, are there actors not involved in at least one of the data practices?
\\
\hline
UC-Goal & Q1. Is the right goal extracted from the scenario?\\

& Q2. Is the extracted UC-Goal, the goal of the UC-User (i.e., primary actor of the scenario) to be accomplished? \\


\hline

UC-DPs & Q1. Are all the data practices extracted in UC-DPs component in the system boundary (i.e., scope)?\\

& Q2. Should the data practice be considered a step?\\

& Q3. Is there any data practice in the extracted UC-DPs that does not contain a flow of personal information?\\

& Q4. Is it clear who is performing the action in the data practice?\\

& Q5. Are there any data practices that are not identified in the extracted UC-DPs?\\
\hline

UC-Steps & Q1. Are all the steps extracted in the extracted UC-Steps component in the system boundary (i.e., scope)?\\

& Q2. Is there any step in the extracted UC-Steps component that does not match the goal or doesn’t help accomplish the goal?\\

& Q3. Is it clear which actor is operating each step? \\

& Q4. Do the steps in the extracted UC-Steps component contain any data practices?\\

& Q5. Are there any steps that are not identified in the extracted UC-Steps?\\





\hline
\end{tabular}
\vspace{0.2cm}
\end{table*}

The quality of UCs and their components greatly affects the quality of other artifacts produced during the development process~\cite{el2006matching}. Low-quality UCs cause defects to propagate to later stages, where the cost of fixing such defects increases significantly~\cite{el2006matching}. 
To analyze the quality of UC components extracted from ChatGPT and address RQ3, we use the guidelines defined by Cockburn~\cite{cockburn2001writing} to create a checklist to inspect the defects in the extracted UCs. Checklist-based inspection techniques are most common in industry~\cite{el2001evaluating}.  Table~\ref{tab:questionnaire} shows this checklist that contains questions related to four general UC components, including (1) \texttt{UC-Actor}, which are the actors involved in the UC (i.e., \texttt{UC-User}, \texttt{UC-System}, and \texttt{UC-ET}), (2) \texttt{UC-Goal}, (3) \texttt{UC-DPs}, and (4) \texttt{UC-Steps}. The quality of \texttt{UC-Name} is not assessed since we believe this component can be extracted from \texttt{UC-Goal}. 

Manually inspecting the UC components extracted for 50 scenarios in the corpus is a time-consuming task. Therefore, we select a sample of 16 scenarios from the corpus. To decide on the sample size, we utilize the number of unique categories in the Apple and Google platforms reported in Table~\ref{tab:appleGoogle_categories}. These categories are representative of the app scenarios in our corpus. Next, we randomly select a scenario from each category, yielding 16 scenarios. The first and last authors manually inspect the extracted UC components in the 16 sampled scenarios to identify potential defects.

\subsection{Exploratory Case Study Design}
We conduct an exploratory case study to address RQ4 and evaluate how changes in ChatGPT prompt affect the quality of the UC components. 
Developers often treat privacy as a secondary concern or a problem for future exploration~\cite{balebako2014improving,li2018coconut}, leading to the construction of systems that fail to provide adequate privacy~\cite{li2022understanding}. 
Therefore, in this study, we focus on ChatGPT's ability to distinguish \texttt{UC-DPs} from \texttt{UC-Steps} and extract them with high quality. \texttt{UC-DPs} contain privacy requirements, and eliciting such requirements from the initial app development life cycle is crucial in privacy-aware software development. 

There are infinite ways to engineer a prompt to improve the quality of components extracted from ChatGPT~\cite{white2023prompt}. We focus on improving the prompt by refining the definitions of \texttt{UC-DPs} and \texttt{UC-Steps} to narrow the search space. In study 2, we use the \textbf{seed prompt} (Figure~\ref{fig:seed-prompt}) with the definitions illustrated in Table~\ref{tab:UCcomponents}. In this study, we engineer two additional prompts (i.e., \textbf{prompt\#1} \& \textbf{prompt\#2}) by refining the definitions of \texttt{UC-DPs} and \texttt{UC-Steps} in the seed prompt. In \textbf{prompt\#1}, we refine the definitions as follows.

\begin{itemize}
    \item \texttt{UC-DPs}: Data practices are specific kinds of interactions between users, systems, or external entities. Data practices convey privacy requirements. A privacy requirement consists of actors with whom the data is shared, actions that are performed on the data, data elements on which actions are performed, and purposes for which data maybe be acted upon. 
\item \texttt{UC-Steps}: A step is an interaction between the user, system, or external entity that is not a data practice. A step is an action the user, system, or external entity performs.
\end{itemize}

In \textbf{prompt\#2}, we further refine prompt\#1 by providing five examples of \texttt{UC-DPs} and \texttt{UC-Steps} from our ground truth for each definition. 

\begin{itemize}
    \item \texttt{UC-DPs}: Data practices are specific kinds of interactions between users, systems, or external entities. Data practices convey privacy requirements. A privacy requirement consists of actors with whom the data is shared, actions that are performed on the data, data elements on which actions are performed, and purposes for which data maybe be acted upon. For example, ``app uses my location,'' ``app collects my height,'' ``user resets password,'' ``user makes purchases on the app,'' ``app uses my name, age, and financial history.''
    \item \texttt{UC-Steps}: A step is an interaction between the user, system, or external entity that is not a data practice. A step is an action the user, system, or external entity performs. For example, ``user opens the Instacart app on their phone,'' user check how many lives are left,'' ``user taps on the safety section at the bottom of the home screen,'' ``user changes sound quality for audio tracks,'' and ``user selects a course to continue.''
\end{itemize}

We ask ChatGPT to extract UC components using prompt\#1 for the 16 sampled scenarios in Section~\ref{sec:evaluation-study3}. We evaluate the quality of the extracted \texttt{UC-DPs} and \texttt{UC-Steps} and inspect the defects using the checklist in Table~\ref{tab:questionnaire}. We also compare prompt\#1 defects with seed prompt defects to see if there is any improvement in the quality of the results. 
We repeat this process for prompt\#2, where we inspect the defects, compare the quality of the extracted \texttt{UC-DPs} and \texttt{UC-Steps} through seed prompt, prompt\#1, and prompt\#2, and report improvements, if any. 


\section{Evaluation Results}\label{sec:results}

In this section, we present our results for studies 1 - 3 as well as our exploratory case study, which address the corresponding research questions (i.e., RQ1-RQ4, respectively). We discuss the results in Section~\ref{sec:discussion}.

\subsection{Study 1 Results}
Study 1 aims to evaluate the quality of the user-authored scenarios and address RQ1. For this reason, we calculate the frequency of scenarios where workers have manually labeled instances of \texttt{UC-Goal}, \texttt{UC-DPs}, and \texttt{UC-Steps}.
Table~\ref{tab:study1Results} presents the frequency of scenarios containing \texttt{UC-Goal}, \texttt{UC-DPs}, and \texttt{UC-Steps} from our scenario corpus. We find that in 47/50 scenarios, the workers can successfully identify goals. Similarly, in 45/50 scenarios, the workers can identify at least one data practice. In all 50 scenarios, the workers can identify at least one step to reach the goal. 

\begin{table}
\centering
\caption{Study 1 Results}
\label{tab:study1Results}
\begin{tabular}{|l|c|c|c|}
\hline
      & \textbf{UC-Goal} & \textbf{UC-DPs} & \textbf{UC-Steps} \\ \hline

\textbf{Frequency}            & 47 & 45 & 50   \\ \hline
\end{tabular}
\end{table}



\subsection{Study 2 Results}
\begin{table}
\centering
\caption{Study 2 Results}
\label{tab:study2Results}
\begin{tabular}{|l|c|p{1.8cm}|p{1.5cm}|c|}
\hline
      & \textbf{EM} & \textbf{F1 without Pre-Processing} & \textbf{F1 with Pre-Processing} & \textbf{SM} \\ \hline

\textbf{UC-Name}            & 0.000     & 0.059  & 0.332 & \textbf{0.459}  \\ \hline
\textbf{UC-Goal}            & 0.000     & 0.256 & 0.285 & \textbf{0.587}  \\ \hline
\textbf{UC-User}           & 0.140     & 0.426 & 0.020 & \textbf{0.468}  \\ \hline
\textbf{UC-System}          & 0.060     & 0.333 & 0.552 & \textbf{0.619}  \\ \hline
\textbf{UC-ET}  & 0.006     & 0.110 & 0.210  & \textbf{0.252}  \\ \hline
\textbf{UC-DPs}   & 0.000     & 0.187 & 0.232  & \textbf{0.475}  \\ \hline
\textbf{UC-Steps}           & 0.000     & 0.357 & 0.436 & \textbf{0.698}  \\ \hline
\end{tabular}
\end{table}


Study 2 aims to evaluate the extent to which ChatGPT can extract UC components from scenarios. To this end, we compare the manually labeled UC components (i.e., ground truth) with the automatically extracted UC components from ChatGPT (i.e., prediction). 
Table~\ref{tab:study2Results} shows the comparison results using three metrics. The EM metric performs poorly when comparing the ground truth with prediction. This metric is the most restrictive that considers the exact match of the strings. 
To mitigate this problem,
we calculate the F1-score based on the precision and recall described in Section~\ref{sec:evaluation}. 
The results show an immediate improvement over the EM metric. 
We also calculate the F1-score after pre-processing the strings in ground truth and prediction. 
We see an improvement in F1-score for all UC components, except for UC-User. We analyze the lists of UC-User components to understand this anomaly. 
We notice that the lists for prediction and ground truth for UC-User have many instances of ``I'' as their element. Since ``I'' is considered a stop word, it gets removed from the prediction and ground truth list during pre-processing. 
Compared to EM and F1-score, the SM metric relies on the context embedded in the strings. Using this metric, we see an improvement across all UC components compared to the other two metrics.

\subsection{Study 3 Results}

In study 3, we assess the quality of the extracted UC components from ChatGPT (i.e., predicted UC components) using a defect checklist (see Section~\ref{sec:evaluation-study3}). 
We report the defects at the scenario level, counting the number of scenarios that did not pass a specific question in the checklist. The inspectors further analyze the reason for each defect by comparing the predicted components with the ground truth. A list of predicted components vs. ground truth for these 16 scenarios is published online\footnote{https://anonymous.4open.science/r/ArtifactsResearch-24FA}. We now summarize our findings based on the checklist. 

\begin{figure*}[]
    \centering
    \includegraphics[width=1.0\textwidth]{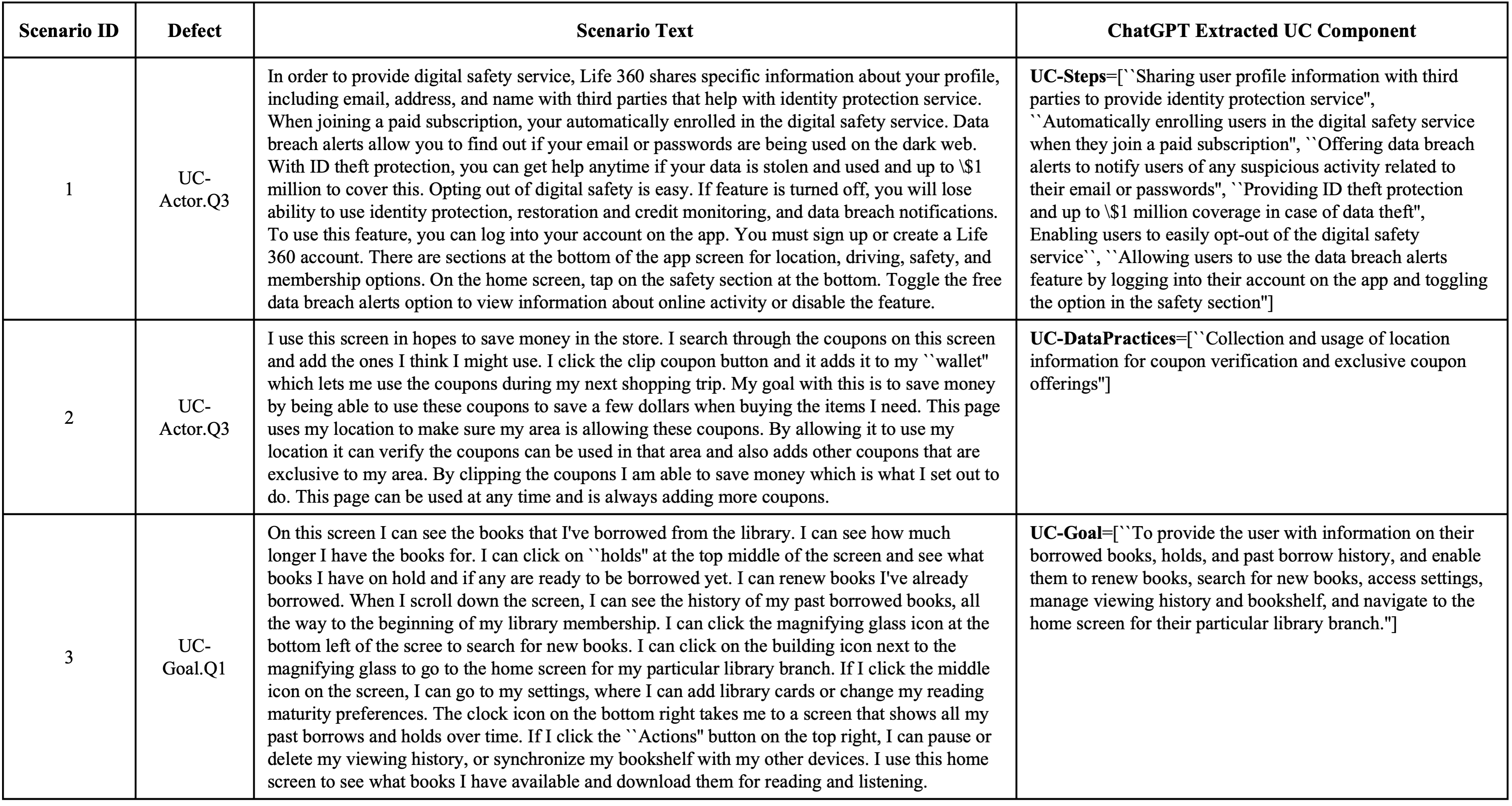}
    \caption{Scenarios and UC Defects Examples}
    \label{fig:scenario-defects-example}
\end{figure*}

\subsubsection{UC-Actor Defects}
The general category of \texttt{UC-Actor} in Table~\ref{tab:questionnaire} entails the UC components associated with \texttt{UC-User}, \texttt{UC-System}, and \texttt{UC-ET}. The following list entails the defects identified in \texttt{UC-Actor}. 
\begin{itemize}
  
  \item Q2: In practice, \texttt{UC-User}, \texttt{UC-System}, and \texttt{UC-ET} are mapped to a simple noun phrase that describes the specific actor (e.g., user, Instacart app, and Paypal). However, in 3/16 scenarios, ChatGPT generates 
  UC components that are sentences describing a specific actor with the actions it performs in the scenario. For example, ChatGPT extracts \texttt{UC-User} as ``The actor who uses the app to make movie purchases'' from a scenario. Further, ChatGPT reports the same result in various ways. For example, we identify \texttt{UC-ET} from 5/16 scenarios where the component contains no elements. However, ChatGPT reports this result as ``None,'' Not Mentioned,'' and ``None Mentioned.''
  
  \item Q3: In 3/16 scenarios, the actors are only involved in \underline{some} of the \texttt{UC-Steps} (e.g., see scenario\#1 in Figure~\ref{fig:scenario-defects-example} and \texttt{UC-Steps} in Table~\ref{tab:Example-QA}).

  \item Q4: In 3/16 scenarios, the actors 
  are only involved in \underline{some} of the \texttt{UC-DPs}. ChatGPT summarizes the actors' interactions and generates new text with a similar context in scenarios (e.g., see scenario\#2, Figure~\ref{fig:scenario-defects-example}). 
\end{itemize}

\subsubsection{UC-Goal Defects} 
\begin{itemize}

  \item Q1: In 3/16 scenarios, the extracted \texttt{UC-Goal} contains \texttt{UC-DPs} \& \texttt{UC-Steps}. Therefore, ChatGPT requires additional information embedded in Heuristics H3 \& H9 in Section~\ref{sec:approach-manual} to extract goals from scenarios (see scenario\#3 in Figure~\ref{fig:scenario-defects-example}). 
  
  
  \item Q2: In 7/16 scenarios, the extracted \texttt{UC-Goal} specifies the goal of the system and not the user. For example, \texttt{UC-Goal}=[``To display a list of movies that the user has purchased''] states the goal of the system/app instead of the user.
\end{itemize}
\subsubsection{UC-DPs Defects} 
\begin{itemize}

  \item Q3: In 2/16 scenarios, the extracted \texttt{UC-DPs} do not contain the information type (i.e., data elements on which actions are performed). For example, a scenario contains data practices, such as ``I enter my weight in the health data,'' and ``App automatically updates when I am on the scale.'' However, ChatGPT extracts three verbs (i.e., ``collection,'' ``sharing,'' and ``usage'') as the data practices. 
  \item Q4: In 12/16 scenarios, the extracted \texttt{UC-DPs} do not contain the actor (see scenario\#2 in Figure~\ref{fig:scenario-defects-example}).  
  \item Q5: We identify 5/16 scenarios where the extracted \texttt{UC-DPs} are missing some data practices when compared with the ground truth. 
  \end{itemize}

\subsubsection{UC-Steps Defects}
\label{UC-Steps Defects}
\begin{itemize}
    \item Q1: In 2/16 scenarios, the extracted \texttt{UC-Steps} contain elements out of the scope/boundary of the system. For example, ``Interact with other bettors and view their bets and reasoning'' and ``Save money by using clipped coupons during shopping trips.''

    \item Q4: In 14/16 scenarios, the extracted \texttt{UC-Steps} do not contain any actor (see scenario\#1 in Figure~\ref{fig:scenario-defects-example}).  

    \item Q4: In 16/16 scenarios, the extracted \texttt{UC-Steps} contain elements that are data practices. For example, ChatGPT extracts the following data practice as a step: ``Sharing user profile information with third parties to provide identity protection service.''  

    \item Q5: We identify 2/16 scenarios where the extracted \texttt{UC-Steps} are missing some steps when compared with the ground truth.
\end{itemize}

\subsection{Exploratory Case Study Results} \label{sec:eval-case-study}
\begin{table}
\centering
\caption{Number of Defects in Exploratory Case Study for 16 Scenarios}
\label{tab:caseStudyResults}
\begin{tabular}{p{0.7cm}ccccc|ccccc}
\hline
Prompt &  \multicolumn{5}{c}{UC-DPs} & \multicolumn{4}{c}{UC-Steps}\\
\hline
 & \textbf{Q1} & \textbf{Q2} & \textbf{Q3} &\textbf{Q4} & \textbf{Q5} &\textbf{Q1} & \textbf{Q2} & \textbf{Q3} & \textbf{Q4} & \textbf{Q5}\\
\hline

Seed & 0 & 0 & 2 & 12 & 5 & 2 & 0 & 14 & 16 & 2\\
\hline
Prompt\#1 & 0  & 0 & 0 & 2 & 7 & 0 & 1 & 6  & 13 & 0\\
\hline
Prompt\#2 & 0  & 0 & 0 & 2 & 5 & 2 & 2 & 5  & 11 & 2\\
\hline
\end{tabular}
\end{table}

We now report the results of our exploratory case study. This study addresses RQ4 and explores if refining ChatGPT prompts can improve the extracted UC components.
Table~\ref{tab:caseStudyResults} illustrates the case study results and compares the number of defects in the scenario level for the extracted \texttt{UC-DPs} and \texttt{UC-Steps} using the seed prompt, prompt\#1, and prompt\#2.
The results reveal that refining the prompt improves the quality of \texttt{UC-DPs} and \texttt{UC-Steps} by adding the actors performing actions. Further, we see an improvement in the quality of \texttt{UC-DPs} due to the flow of information types being mentioned. In addition, the quality of \texttt{UC-Steps} gradually increases through prompt refinement. The number of scenarios where the \texttt{UC-Steps} components contain data practices decreases from 16 to 11 when using prompt\#2. We also check whether the missing data practices from \texttt{UC-Steps} are actually presented in the \texttt{UC-DPs} for the 11 scenarios. Our analysis shows that 3/11 scenarios now entail \texttt{UC-DPs} and \texttt{UC-Steps} with unique elements as interactions. 

\section{Discussion}\label{sec:discussion}

We now review the research questions in the context of the results. 
RQ1 checks the quality of user-authored scenarios considering the goal, steps \& information types required to achieve the goal. We identify 47 \& 50 goals and steps, respectively. However, only 45 scenarios contain data practices for the following reasons: (1) during manual labeling, the workers can agree that the scenario does not contain sensitive information; or (2) the scenario author has not provided any sensitive information required to achieve the goal.

RQ2 assesses the extent of ChatGPT's ability to extract UC components from scenarios. In study 2, we compare the ground truth UC components with the extracted components from ChatGPT.
The EM metric is more strict in comparing characters than the F1-score, which is evident by an average similarity value of 0.029 and 0.276, respectively. Pre-processing text before calculating the F1-score improves the average similarity to 0.2995 over seven UC components. 
The results show an improvement with an average of 0.50 similarity across all UC components when using the SM metric. This metric outperforms string comparison-based metrics. Even using SM, the results reveal low similarity for some components, including UC-ET and UC-DPs. Without further qualitative analysis, we are unable to identify the reasons behind the discrepancy. Therefore, we design study 3 to address RQ3 and identify the defects in the extracted UC components in a qualitative manner. 
Through our analysis, ChatGPT lacks domain knowledge to extract \texttt{UC-DPs} that contain privacy requirements. The workers utilize such domain knowledge when labeling the scenarios manually, as they all have backgrounds and experience in privacy research. Further, during the manual labeling task, the workers identify heuristics that help clarify boundary and edge cases. However, ChatGPT fails to extract quality UC components when lacking such heuristics. Finally, ChatGPT is a LLM capable of summarizing text. We often observe that ChatGPT summarizes multiple steps or data practices into one sentence. This potentially has unwanted effects as data practices (i.e., privacy requirements) are summarized and presented in an abstract and ambiguous manner, increasing the privacy risk associated with such practices~\cite{bhatia2016theory,hosseini2021RE}.
We also see the effect of text generation in extracting the \texttt{UC-Goal}. In practice, the \texttt{UC-Goal} should be narrated as the primary user's goal. However, in many instances, ChatGPT generates a new text which describes the app's goal in the interactions. 

To address RQ4 and explore the impact of prompt engineering on the quality of ChatGPT's results, we conduct a case study on the sampled 16 scenarios. We compare the quality of \texttt{UC-DPs} and \texttt{UC-Steps} extracted using three prompts: seed, prompt\#1, and prompt\#2. Prompt\#1 results show that in 6/16 scenarios, the extracted UC-DPs are in a dictionary format where the keys are ``Data Sharing'', ``Data Action'', ``Data Element'', and ``Data Purpose''. 
We do not consider this format a defect since the quality checklist (see Table~\ref{tab:questionnaire}) does not contain questions regarding the presentation format. However, the extracted UC components' format can be improved by engineering a prompt containing a specific template for the results~\cite{white2023prompt} or by providing examples of data practices and steps. After refining prompt\#1 by adding examples of data practices and steps, we observe an improvement in the UC component presentation format. 
Overall, the results of this case study reveal that the number of defects either remains stable or decreases by refining the prompt.

\section{Threats to Validity}\label{sec:threats}
\noindent \textbf{Internal Validity} concerns whether the inferences drawn from the studies are valid. 
Evaluation of the UC components extracted by ChatGPT are based on a limited set of components. Components such as pre- and post-conditions, and alternative scenarios are not considered. However, we regard the chosen components to be representative of the standard definition of a UC\cite{cockburn2001writing}, and those secondary components that are not included would have minimal effect on the outcome of the study. In addition, we design the quality checklist based on the chosen components and the list of questions and defects available in the literature~\cite{cockburn2001writing,anda2002towards, el2012constructing}.
The quality of the prompt in ChatGPT can directly affect the quality of the generated output. For example, asking the same question with a different grammatical structure and word order \& selection could change the generated output. 

\noindent\textbf{External Validity} concerns the extent to which the results generalize beyond the experimental setting. 
Even with controlled and consistent prompts, ChatGPT's output is non-deterministic. Because of the generative goals of LLMs, it is often desirable for them to produce inconsistent output. As a result, given the same scenarios, different UC components may be produced. While this can affect the consistency in quality of the generated UCs, we find the output to vary mostly in wording with very little variance to the actual components produced. 
Further, ChatGPT is a continually-evolving model. As such, the quality of UC components it produces is likely to improve as new versions are released. 

\noindent\textbf{Reliability} indicates that the researchers' approach is consistent across different researchers and different projects~\cite{gibbs2018}. To ensure the reliability of our research, we measure and report the inter-coder agreement~\cite{creswell2017} using Fleiss' Kappa~\cite{fleiss1971measuring} for the UC component labeling task. 

\section{Conclusion \& Future Work} \label{sec:conclusion}
In this paper, we utilize scenarios that enable small app-developing companies to identify a sustainable and low-cost source of requirements. 
We also evaluate ChatGPT's ability to extract UC components from scenarios to further assist developers. Our results reveal that ChatGPT requires additional domain knowledge to extract the UC components. Further, we explore whether refining ChatGPT's prompt with domain knowledge can help improve the quality of the extracted components. We observe that such refinements improve the quality of the components in our sample set. In future work, we envision (1) methods to fine-tune large language models (LLMs) with domain knowledge; (2) identifying prompt patterns to improve the quality of automated use case extraction from LLMs; (3) exploring the ability of LLMs in generate object-oriented design artifacts from textual requirements.


\bibliographystyle{IEEEtran}
\bibliography{references,ml,policies,domainModel}

\begin{thebibliography}{10}
\providecommand{\url}[1]{#1}
\csname url@samestyle\endcsname
\providecommand{\newblock}{\relax}
\providecommand{\bibinfo}[2]{#2}
\providecommand{\BIBentrySTDinterwordspacing}{\spaceskip=0pt\relax}
\providecommand{\BIBentryALTinterwordstretchfactor}{4}
\providecommand{\BIBentryALTinterwordspacing}{\spaceskip=\fontdimen2\font plus
\BIBentryALTinterwordstretchfactor\fontdimen3\font minus \fontdimen4\font\relax}
\providecommand{\BIBforeignlanguage}[2]{{%
\expandafter\ifx\csname l@#1\endcsname\relax
\typeout{** WARNING: IEEEtran.bst: No hyphenation pattern has been}%
\typeout{** loaded for the language `#1'. Using the pattern for}%
\typeout{** the default language instead.}%
\else
\language=\csname l@#1\endcsname
\fi
#2}}
\providecommand{\BIBdecl}{\relax}
\BIBdecl

\bibitem{wasserman2012software}
A.~I. Wasserman and P.~Freeman, \emph{Software Engineering Education: Needs and Objectives Proceedings of an Interface Workshop}.\hskip 1em plus 0.5em minus 0.4em\relax Springer Science \& Business Media, 2012.

\bibitem{abad2017learn}
Z.~S.~H. Abad, S.~D. Sims, A.~Cheema, M.~B. Nasir, and P.~Harisinghani, ``Learn more, pay less! lessons learned from applying the wizard-of-oz technique for exploring mobile app requirements,'' in \emph{2017 IEEE 25th RE Workshops (REW)}.\hskip 1em plus 0.5em minus 0.4em\relax IEEE, 2017, pp. 132--138.

\bibitem{oriol2018fame}
M.~Oriol, M.~Stade, F.~Fotrousi, S.~Nadal, J.~Varga, N.~Seyff, A.~Abello, X.~Franch, J.~Marco, and O.~Schmidt, ``Fame: supporting continuous requirements elicitation by combining user feedback and monitoring,'' in \emph{2018 ieee 26th RE}.\hskip 1em plus 0.5em minus 0.4em\relax IEEE, 2018, pp. 217--227.

\bibitem{wei2022towards}
J.~Wei, A.-L. Courbis, T.~Lambolais, P.~L. Bernard, and G.~Dray, ``Towards boosting requirements engineering of a health monitoring app by analysing similar apps: A vision paper,'' in \emph{2022 IEEE 30th RE Workshops (REW)}.\hskip 1em plus 0.5em minus 0.4em\relax IEEE, 2022, pp. 75--80.

\bibitem{dalpiaz2019re}
F.~Dalpiaz and M.~Parente, ``Re-swot: from user feedback to requirements via competitor analysis,'' in \emph{International working conference on requirements engineering: foundation for software quality}.\hskip 1em plus 0.5em minus 0.4em\relax Springer, 2019, pp. 55--70.

\bibitem{di2016would}
A.~Di~Sorbo, S.~Panichella, C.~V. Alexandru, J.~Shimagaki, C.~A. Visaggio, G.~Canfora, and H.~C. Gall, ``What would users change in my app? summarizing app reviews for recommending software changes,'' in \emph{Proceedings of the 2016 24th ACM SIGSOFT International Symposium on Foundations of Software Engineering}, 2016, pp. 499--510.

\bibitem{maalej2015bug}
W.~Maalej and H.~Nabil, ``Bug report, feature request, or simply praise? on automatically classifying app reviews,'' in \emph{2015 IEEE 23rd RE}.\hskip 1em plus 0.5em minus 0.4em\relax IEEE, 2015, pp. 116--125.

\bibitem{genc2017systematic}
N.~Genc-Nayebi and A.~Abran, ``A systematic literature review: Opinion mining studies from mobile app store user reviews,'' \emph{Journal of Systems and Software}, vol. 125, pp. 207--219, 2017.

\bibitem{chen2014ar}
N.~Chen, J.~Lin, S.~C. Hoi, X.~Xiao, and B.~Zhang, ``Ar-miner: mining informative reviews for developers from mobile app marketplace,'' in \emph{Proceedings of the 36th international conference on software engineering}, 2014, pp. 767--778.

\bibitem{rago2013uncovering}
A.~Rago, C.~Marcos, and J.~A. Diaz-Pace, ``Uncovering quality-attribute concerns in use case specifications via early aspect mining,'' \emph{Requirements Engineering}, vol.~18, pp. 67--84, 2013.

\bibitem{luisa2004market}
M.~Luisa, F.~Mariangela, and N.~I. Pierluigi, ``Market research for requirements analysis using linguistic tools,'' \emph{Requirements Engineering}, vol.~9, pp. 40--56, 2004.

\bibitem{cockburn2001writing}
A.~Cockburn, \emph{Writing effective use cases}.\hskip 1em plus 0.5em minus 0.4em\relax Pearson Education India, 2001.

\bibitem{fantechi2003applications}
A.~Fantechi, S.~Gnesi, G.~Lami, and A.~Maccari, ``Applications of linguistic techniques for use case analysis,'' \emph{Requirements Engineering}, vol.~8, pp. 161--170, 2003.

\bibitem{el2012constructing}
M.~El-Attar and J.~Miller, ``Constructing high quality use case models: a systematic review of current practices,'' \emph{Requirements Engineering}, vol.~17, pp. 187--201, 2012.

\bibitem{al2018use}
A.~Al-Hroob, A.~T. Imam, and R.~Al-Heisa, ``The use of artificial neural networks for extracting actions and actors from requirements document,'' \emph{Information and Software Technology}, vol. 101, pp. 1--15, 2018.

\bibitem{anda2001quality}
B.~Anda, D.~Sj{\o}berg, and M.~J{\o}rgensen, ``Quality and understandability of use case models,'' in \emph{ECOOP 2001—Object-Oriented Programming: 15th European Conference Budapest, Hungary, June 18--22, 2001 Proceedings 15}.\hskip 1em plus 0.5em minus 0.4em\relax Springer, 2001, pp. 402--428.

\bibitem{el2006matching}
M.~El-Attar and J.~Miller, ``Matching antipatterns to improve the quality of use case models,'' in \emph{14th IEEE RE}.\hskip 1em plus 0.5em minus 0.4em\relax IEEE, 2006, pp. 99--108.

\bibitem{el2006agaduc}
------, ``{AGADUC}: Towards a more precise presentation of functional requirement in use case mod,'' in \emph{Fourth International Conference on Software Engineering Research, Management and Applications (SERA'06)}.\hskip 1em plus 0.5em minus 0.4em\relax IEEE, 2006, pp. 346--353.

\bibitem{el2010improving}
------, ``Improving the quality of use case models using antipatterns,'' \emph{Software \& systems modeling}, vol.~9, pp. 141--160, 2010.

\bibitem{vaswani2017attention}
A.~Vaswani, N.~Shazeer, N.~Parmar, J.~Uszkoreit, L.~Jones, A.~N. Gomez, {\L}.~Kaiser, and I.~Polosukhin, ``Attention is all you need,'' \emph{Advances in neural information processing systems}, vol.~30, 2017.

\bibitem{white2023prompt}
J.~White, Q.~Fu, S.~Hays, M.~Sandborn, C.~Olea, H.~Gilbert, A.~Elnashar, J.~Spencer-Smith, and D.~C. Schmidt, ``A prompt pattern catalog to enhance prompt engineering with chatgpt,'' \emph{arXiv preprint arXiv:2302.11382}, 2023.

\bibitem{yang2015identification}
H.~Yang and P.~Liang, ``Identification and classification of requirements from app user reviews.'' in \emph{SEKE}, 2015, pp. 7--12.

\bibitem{pagano2013user}
D.~Pagano and W.~Maalej, ``User feedback in the appstore: An empirical study,'' in \emph{2013 21st IEEE RE}.\hskip 1em plus 0.5em minus 0.4em\relax IEEE, 2013, pp. 125--134.

\bibitem{panichella2015can}
S.~Panichella, A.~Di~Sorbo, E.~Guzman, C.~A. Visaggio, G.~Canfora, and H.~C. Gall, ``How can i improve my app? classifying user reviews for software maintenance and evolution,'' in \emph{2015 IEEE international conference on software maintenance and evolution (ICSME)}.\hskip 1em plus 0.5em minus 0.4em\relax IEEE, 2015, pp. 281--290.

\bibitem{harman2012app}
M.~Harman, Y.~Jia, and Y.~Zhang, ``App store mining and analysis: Msr for app stores,'' in \emph{2012 9th IEEE working conference on mining software repositories (MSR)}.\hskip 1em plus 0.5em minus 0.4em\relax IEEE, 2012, pp. 108--111.

\bibitem{iacob2013retrieving}
C.~Iacob and R.~Harrison, ``Retrieving and analyzing mobile apps feature requests from online reviews,'' in \emph{2013 10th working conference on mining software repositories (MSR)}.\hskip 1em plus 0.5em minus 0.4em\relax IEEE, 2013, pp. 41--44.

\bibitem{dkabrowski2020mining}
J.~D{\k{a}}browski, E.~Letier, A.~Perini, and A.~Susi, ``Mining user opinions to support requirement engineering: an empirical study,'' in \emph{Advanced Information Systems Engineering: 32nd International Conference, CAiSE 2020, Grenoble, France, June 8--12, 2020, Proceedings 32}.\hskip 1em plus 0.5em minus 0.4em\relax Springer, 2020, pp. 401--416.

\bibitem{sutcliffe1998scenario}
A.~Sutcliffe, ``Scenario-based requirements analysis,'' \emph{Requirements engineering}, vol.~3, pp. 48--65, 1998.

\bibitem{alexander2005scenarios}
I.~F. Alexander and N.~Maiden, \emph{Scenarios, stories, use cases: through the systems development life-cycle}.\hskip 1em plus 0.5em minus 0.4em\relax John Wiley \& Sons, 2005.

\bibitem{anton1998representational}
A.~I. Ant{\'o}n and C.~Potts, ``A representational framework for scenarios of system use,'' \emph{Requirements Engineering}, vol.~3, pp. 219--241, 1998.

\bibitem{weidenhaupt1998scenarios}
K.~Weidenhaupt, K.~Pohl, M.~Jarke, and P.~Haumer, ``Scenarios in system development: current practice,'' \emph{IEEE software}, vol.~15, no.~2, pp. 34--45, 1998.

\bibitem{hibshi2021systemic}
H.~Hibshi, S.~T. Jones, and T.~D. Breaux, ``A systemic approach for natural language scenario elicitation of security requirements,'' \emph{IEEE Transactions on Dependable and Secure Computing}, vol.~19, no.~6, pp. 3579--3591, 2021.

\bibitem{shen2022domain}
Y.~Shen and T.~Breaux, ``Domain model extraction from user-authored scenarios and word embeddings,'' in \emph{2022 IEEE 30th RE Workshops (REW)}.\hskip 1em plus 0.5em minus 0.4em\relax IEEE, 2022, pp. 143--151.

\bibitem{potts1994inquiry}
C.~Potts, K.~Takahashi, and A.~I. Anton, ``Inquiry-based requirements analysis,'' \emph{IEEE software}, vol.~11, no.~2, pp. 21--32, 1994.

\bibitem{johnson1988computer}
P.~N. Johnson-Laird, \emph{The computer and the mind: An introduction to cognitive science}.\hskip 1em plus 0.5em minus 0.4em\relax Harvard University Press, 1988.

\bibitem{jacobson1995use}
I.~Jacobson, ``The use-case construct in object-oriented software engineering,'' in \emph{Scenario-based design: envisioning work and technology in system development}, 1995, pp. 309--336.

\bibitem{larman1998applying}
C.~Larman, \emph{Applying UML and patterns}.\hskip 1em plus 0.5em minus 0.4em\relax Prentice Hall Englewood Cliffs, NJ, 1998, vol.~2.

\bibitem{insfran2002requirements}
E.~Insfr{\'a}n, O.~Pastor, and R.~Wieringa, ``Requirements engineering-based conceptual modelling,'' \emph{Requirements Engineering}, vol.~7, no.~2, p.~61, 2002.

\bibitem{bittner2003use}
K.~Bittner and I.~Spence, \emph{Use case modeling}.\hskip 1em plus 0.5em minus 0.4em\relax Addison-Wesley Professional, 2003.

\bibitem{ouergaard2005use}
G.~K.~P. {\"O}uergaard, ``Use gases patterns and blueprints,'' 2005.

\bibitem{anda2002towards}
B.~Anda and D.~I. Sj{\o}berg, ``Towards an inspection technique for use case models,'' in \emph{Proceedings of the 14th international conference on Software engineering and knowledge engineering}, 2002, pp. 127--134.

\bibitem{kundi2017use}
M.~Kundi and R.~Chitchyan, ``Use case elicitation with framenet frames,'' in \emph{2017 IEEE 25th RE Workshops (REW)}.\hskip 1em plus 0.5em minus 0.4em\relax IEEE, 2017, pp. 224--231.

\bibitem{dalpiaz2020conceptualizing}
F.~Dalpiaz and A.~Sturm, ``Conceptualizing requirements using user stories and use cases: a controlled experiment,'' in \emph{Requirements Engineering: Foundation for Software Quality: 26th International Working Conference, REFSQ 2020, Pisa, Italy, March 24--27, 2020, Proceedings 26}.\hskip 1em plus 0.5em minus 0.4em\relax Springer, 2020, pp. 221--238.

\bibitem{maiden2004model}
N.~A. Maiden, S.~V. Jones, S.~Manning, J.~Greenwood, and L.~Renou, ``Model-driven requirements engineering: synchronising models in an air traffic management case study,'' in \emph{Advanced Information Systems Engineering: 16th International Conference, CAiSE 2004, Riga, Latvia, June 7-11, 2004. Proceedings 16}.\hskip 1em plus 0.5em minus 0.4em\relax Springer, 2004, pp. 368--383.

\bibitem{yue2011systematic}
T.~Yue, L.~C. Briand, and Y.~Labiche, ``A systematic review of transformation approaches between user requirements and analysis models,'' \emph{Requirements engineering}, vol.~16, pp. 75--99, 2011.

\bibitem{devlin2018bert}
J.~Devlin, M.-W. Chang, K.~Lee, and K.~Toutanova, ``Bert: Pre-training of deep bidirectional transformers for language understanding,'' \emph{arXiv preprint arXiv:1810.04805}, 2018.

\bibitem{liu2019roberta}
Y.~Liu, M.~Ott, N.~Goyal, J.~Du, M.~Joshi, D.~Chen, O.~Levy, M.~Lewis, L.~Zettlemoyer, and V.~Stoyanov, ``Roberta: A robustly optimized bert pretraining approach,'' \emph{arXiv preprint arXiv:1907.11692}, 2019.

\bibitem{radford2018improving}
A.~Radford, K.~Narasimhan, T.~Salimans, I.~Sutskever \emph{et~al.}, ``Improving language understanding by generative pre-training,'' 2018.

\bibitem{radford2019language}
A.~Radford, J.~Wu, R.~Child, D.~Luan, D.~Amodei, I.~Sutskever \emph{et~al.}, ``Language models are unsupervised multitask learners,'' \emph{OpenAI blog}, vol.~1, no.~8, p.~9, 2019.

\bibitem{brown2020language}
T.~Brown, B.~Mann, N.~Ryder, M.~Subbiah, J.~D. Kaplan, P.~Dhariwal, A.~Neelakantan, P.~Shyam, G.~Sastry, A.~Askell \emph{et~al.}, ``Language models are few-shot learners,'' \emph{Advances in neural information processing systems}, vol.~33, pp. 1877--1901, 2020.

\bibitem{ouyang2022training}
L.~Ouyang, J.~Wu, X.~Jiang, D.~Almeida, C.~L. Wainwright, P.~Mishkin, C.~Zhang, S.~Agarwal, K.~Slama, A.~Ray \emph{et~al.}, ``Training language models to follow instructions with human feedback,'' \emph{arXiv preprint arXiv:2203.02155}, 2022.

\bibitem{sobania2023analysis}
D.~Sobania, M.~Briesch, C.~Hanna, and J.~Petke, ``An analysis of the automatic bug fixing performance of chatgpt,'' \emph{arXiv preprint arXiv:2301.08653}, 2023.

\bibitem{zhou2022large}
Y.~Zhou, A.~I. Muresanu, Z.~Han, K.~Paster, S.~Pitis, H.~Chan, and J.~Ba, ``Large language models are human-level prompt engineers,'' \emph{arXiv preprint arXiv:2211.01910}, 2022.

\bibitem{liu2022design}
V.~Liu and L.~B. Chilton, ``Design guidelines for prompt engineering text-to-image generative models,'' in \emph{Proceedings of the 2022 CHI}, 2022, pp. 1--23.

\bibitem{breaux2014eddy}
T.~D. Breaux, H.~Hibshi, and A.~Rao, ``Eddy, a formal language for specifying and analyzing data flow specifications for conflicting privacy requirements,'' \emph{Requirements Engineering}, vol.~19, no.~3, pp. 281--307, 2014.

\bibitem{hoepman2014privacy}
J.-H. Hoepman, ``Privacy design strategies,'' in \emph{ICT Systems Security and Privacy Protection: 29th IFIP TC 11 International Conference, SEC 2014, Marrakech, Morocco, June 2-4, 2014. Proceedings 29}.\hskip 1em plus 0.5em minus 0.4em\relax Springer, 2014, pp. 446--459.

\bibitem{colesky2016critical}
M.~Colesky, J.-H. Hoepman, and C.~Hillen, ``A critical analysis of privacy design strategies,'' in \emph{2016 IEEE security and privacy workshops (SPW)}.\hskip 1em plus 0.5em minus 0.4em\relax IEEE, 2016, pp. 33--40.

\bibitem{colesky2016privacy}
M.~Colesky and S.~Ghanavati, ``Privacy shielding by design—a strategies case for near-compliance,'' in \emph{2016 IEEE 24th RE Workshops (REW)}.\hskip 1em plus 0.5em minus 0.4em\relax IEEE, 2016, pp. 271--275.

\bibitem{fleiss1971measuring}
J.~L. Fleiss, ``Measuring nominal scale agreement among many raters.'' \emph{Psychological bulletin}, vol.~76, no.~5, p. 378, 1971.

\bibitem{landis1977measurement}
J.~R. Landis and G.~G. Koch, ``The measurement of observer agreement for categorical data,'' \emph{biometrics}, pp. 159--174, 1977.

\bibitem{risch-etal-2021-semantic}
\BIBentryALTinterwordspacing
J.~Risch, T.~M{\"o}ller, J.~Gutsch, and M.~Pietsch, ``Semantic answer similarity for evaluating question answering models,'' in \emph{Proceedings of the 3rd Workshop on Machine Reading for Question Answering}.\hskip 1em plus 0.5em minus 0.4em\relax Punta Cana, Dominican Republic: Association for Computational Linguistics, Nov. 2021, pp. 149--157. [Online]. Available: \url{https://aclanthology.org/2021.mrqa-1.15}
\BIBentrySTDinterwordspacing

\bibitem{el2001evaluating}
K.~El~Emam and O.~Laitenberger, ``Evaluating capture-recapture models with two inspectors,'' \emph{IEEE Transactions on Software Engineering}, vol.~27, no.~9, pp. 851--864, 2001.

\bibitem{balebako2014improving}
R.~Balebako and L.~Cranor, ``Improving app privacy: Nudging app developers to protect user privacy,'' \emph{IEEE Security \& Privacy}, vol.~12, no.~4, pp. 55--58, 2014.

\bibitem{li2018coconut}
T.~Li, Y.~Agarwal, and J.~I. Hong, ``Coconut: An ide plugin for developing privacy-friendly apps,'' \emph{Proceedings of the ACM on Interactive, Mobile, Wearable and Ubiquitous Technologies}, vol.~2, no.~4, pp. 1--35, 2018.

\bibitem{li2022understanding}
\BIBentryALTinterwordspacing
T.~Li, K.~Reiman, Y.~Agarwal, L.~F. Cranor, and J.~I. Hong, ``\BIBforeignlanguage{en}{Understanding challenges for developers to create accurate privacy nutrition labels},'' in \emph{\BIBforeignlanguage{en}{CHI}}.\hskip 1em plus 0.5em minus 0.4em\relax New Orleans LA USA: ACM, Apr 2022, p. 1–24. [Online]. Available: \url{https://dl.acm.org/doi/10.1145/3491102.3502012}
\BIBentrySTDinterwordspacing

\bibitem{bhatia2016theory}
J.~Bhatia, T.~D. Breaux, J.~R. Reidenberg, and T.~B. Norton, ``A theory of vagueness and privacy risk perception,'' in \emph{2016 IEEE 24th RE}.\hskip 1em plus 0.5em minus 0.4em\relax IEEE, 2016, pp. 26--35.

\bibitem{hosseini2021RE}
M.~B. Hosseini, J.~Heaps, R.~Slavin, T.~Breaux, and J.~Niu, ``Ambiguity and generality in natural language privacy policies,'' in \emph{2021 29th IEEE International Requirements Engineering Conference (RE)}.\hskip 1em plus 0.5em minus 0.4em\relax IEEE, 2021.

\bibitem{gibbs2018}
G.~R. Gibbs, \emph{Analyzing qualitative data}.\hskip 1em plus 0.5em minus 0.4em\relax Sage, 2018, vol.~6.

\bibitem{creswell2017}
J.~W. Creswell and J.~D. Creswell, \emph{Research design: Qualitative, quantitative, and mixed methods approaches}.\hskip 1em plus 0.5em minus 0.4em\relax Sage publications, 2017.

\end{thebibliography}

\end{document}